\documentclass[a4paper]{report}
\usepackage[utf8]{inputenc}
\usepackage[T1]{fontenc}
\usepackage{RJournal}
\usepackage{amsmath,amssymb,array}
\usepackage{booktabs}
\usepackage{subcaption}

\graphicspath{{./figures/}}

\begin{document}

\sectionhead{Contributed research article}
\volume{14}
\volnumber{2}
\year{2022}
\month{June}

\begin{article}

\title{R-miss-tastic: a unified platform for missing values methods and workflows}
\author{by Imke Mayer, Aude Sportisse, Nicholas Tierney, Nathalie Vialaneix and Julie Josse}

\maketitle

\abstract{
Missing values are unavoidable when working with data. 
To help address this challenge, we have launched the \samp{R-miss-tastic} platform, which aims to provide an overview of standard missing values problems, methods, and relevant implementations of methodologies. Beyond gathering and organizing a large majority of the material on missing data (bibliography, courses, tutorials, implementations), \samp{R-miss-tastic} covers the development of standardized analysis workﬂows. Indeed, we have developed several pipelines in {R} and {Python} to allow for hands-on illustration of and recommendations on missing values handling in various statistical tasks such as matrix completion, estimation, and prediction, while ensuring reproducibility of the analyses. Finally, the platform is dedicated to users who analyze incomplete data, researchers who want to compare their methods and search for an up-to-date bibliography, and also teachers who are looking for didactic materials (notebooks, video, slides). 
}

\section{Context and motivation}

\label{sec:intro_rmiss}

Missing data are unavoidable as soon as collecting or acquiring data is 
involved. They occur for many reasons including: individuals choosing not to 
answer survey questions, measurement devices failing, or data having simply not been recorded.
Their presence becomes even more important as data are now
obtained at increasing velocity and volume, and from heterogeneous sources not originally designed
to be analyzed together. As pointed out by \citet{zhu2019high}, \samp{one of the
ironies of working with Big Data is that missing data play an ever more
significant role, and often present serious difficulties for analysis}.
Despite this, the approach most commonly implemented by default in
software is to toss out cases with missing values. At best, this is inefficient
because it wastes information from the partially observed cases. At worst, it
results in biased estimates, particularly when the distributions of the missing
values are systematically different from those of the observed values \citep[e.g.,][Chap. 2]{enders_AMDA2010}.

However, handling missing data in a more efficient and relevant way (than
limiting the analysis to solely the complete cases) has attracted a lot of
attention in the literature in the last two decades. In particular, a number of
reference books have been published
\citep{Schafer2002,VanBuuren2012,Carpenter2012,little2019statistical} and the topic is an
active field of research \citep{josse_reiter_2018}. The diversity of the 
missing data problems means there is great variety in the proposed and studied methods. 
They include model-based approaches, integrating
likelihoods or posterior distributions over missing values, filling in missing values in a realistic way with single, or multiple
imputations, or weighting of observations, 
appealing to ideas from the design-based literature in survey
sampling.
The multiplicity of the available solutions makes sense because there is no single
solution or tool to manage missing data: the appropriate methodology to handle
them depends on many features, such as the objective of the analysis, type of data, the type of
missing data and their pattern.
Some of these methods are
available in various software solutions. As R \citep{R} is one of the
main pieces of software for statisticians and data scientists, with its development starting almost three decades ago \citep{ihaka_1998}, R offers the largest number of implemented approaches. This is also due to
its ease in incorporating new methods and its modular packaging system.
Currently, there are over 270 {R} packages on CRAN that mention missing data
or imputation in their DESCRIPTION files. These packages serve many different
applications, data types or types of analysis. More precisely, exploratory
and visualization tools for missing data are available in packages like
\pkg{naniar}, \pkg{VIM}, and \pkg{MissingDataGUI} \citep{naniar,Tierney2018,VIM,Cheng2015}.
Imputation methods are included in packages like \pkg{mice}, \pkg{Amelia}, and
\pkg{mi} \citep{mice,amelia,mi}. Other packages focus on dealing with complex,
heterogeneous (categorical, quantitative, ordinal variables) data or with large
dimension multi-level data, such as \pkg{missMDA}, and \pkg{MixedDataImpute}
\citep{josse2016missmda,Murray2015}. 
Besides {R}, other languages such as {Python} \citep{python}, which currently only have few publicly available implementations of methods that handle missing values, offer increasingly more  solutions. Two prominent  examples are: 1) the \pkg{scikit-learn} library \citep{pedregosa2011scikit} which has recently added a module for missing values imputation; and 2) the \pkg{DataWig} library \citep{datawig} which provides a framework to learn to impute incomplete data tables.

Despite the large range of options, 
missing data are often not handled appropriately by practitioners. This may be for several reasons. First, the plethora of options can be a double-edged sword; while it is great to have many options, finding the most appropriate method is challenging as there are so many. Second, the topic of missing data is often itself missing from many 
statistics and data science syllabuses, despite its omnipresence in 
data. So, when faced with missing data, practitioners are left powerless; quite 
possibly never having been taught about missing data, they do not know how to approach the problem, the dangers of  mismanagement, how to navigate the methods, software, or how to choose 
the most appropriate method or workflow.

To help promote better management and understanding of missing data, we have released \samp{R-miss-tastic}, an open platform for missing values. 
The platform takes the form of a reference website\footnote{\url{https://rmisstastic.netlify.com/}}, which collects, organizes and produces
material on missing data. 
It has been conceived by an infrastructure steering committee
 working group (ISC; its members are authors of this article), which first provided a CRAN Task View\footnote{\url{https://CRAN.R-project.org/package=ctv}} on missing data\footnote{\url{https://cran.r-project.org/web/views/MissingData.html}} that
lists and organizes existing {R} packages on the topic.
The \samp{R-miss-tastic} platform extends and builds on the CRAN Task View by collecting, creating and organizing articles, tutorials,
documentation, and  workflows for analyses with missing data. 

This platform is easily extendable and well documented, allowing it to seamlessly incorporate future works and research in missing values. 
The intent of the platform is to foster a welcoming community, within and beyond the {R} community.
\samp{R-miss-tastic} has been designed to be accessible for a wide audience with different levels of prior knowledge, needs, and
questions. This includes students, teachers, statisticians, and researchers. Students can use it as complementary course materials. Teachers can use it as a reference website for their own classes. Statisticians and researchers can find example analysis workflows, or even contribute information for specific areas and find collaborators.

The platform provides
new tutorials, examples and pipelines of analyses that we have developed with missing data spanning the entirety of an analysis. These have been developed in {R} and in {Python}, implementing standard methods for generating missing values, and for analyzing them under different perspectives. In addition, we reference publicly available datasets that are commonly used as benchmark for new missing values methodologies. The developed pipelines cover the entirety of a data analysis: exploratory analyses, establishing statistical and machine learning models, analysis diagnostics, and finally  interpreting results obtained from incomplete data. We hope these pipelines also serve as a guide when choosing a method to handle missing values.

The remainder of the article is organized as follows:
In the section entitled ``Structure and content of the platform'' we
describe the different components of the platform, the structure that has been chosen, and the target audience. The section is organized as the platform
itself, starting by describing materials for less advanced users then materials
for researchers and finally resources for practical implementation. We then detail the implementation and use-cases of the provided {R} and  {Python} workflows in the following section entitled ``Details of the missing values workflows''. Finally, in the conclusion, we outline an overview of planed future developments
for the platform and interesting areas in missing values research that we would like to bring to a wider audience.

\section{Structure and content of the platform}
\label{sec:struct}

The \samp{R-miss-tastic} platform is released at
\url{https://rmisstastic.netlify.com/}. It has been developed using the {R}
package \pkg{blogdown} \citep{xie_etal_bCWRM2017} which generates static websites using Hugo\footnote{\url{https://gohugo.io/}}. Live examples have been included using
the tool \url{https://rdrr.io/snippets/} provided by the website \samp{R Package
Documentation}. The source code and materials of the platform have been made publicly available on
GitHub at \url{https://github.com/R-miss-tastic}, which provides a transparent record of the platform's development, and facilitates contributions from the community.

We now discuss the structure of the \samp{R-miss-tastic} platform, the aim and content of each subsection, and highlight key features of the platform.

\subsection{Missing values workflows}

An important contribution and novelty of this work is the proposal of several workflows that allow for a hands-on illustration of classical analyses with missing values, both on simulated data and on publicly available real-world data. These workflows are provided both in {R} and in {Python} code and cover the following topics:
\begin{itemize}
\item \textit{How to generate missing values?} Generate missing values under
different mechanisms, on complete or incomplete datasets. This is useful when performing simulations
to compare methods that impute or handle missing data.
\item \textit{How to do statistical inference with missing values?} In particular, we focus on different solutions for estimating
linear and logistic regression parameters with missing covariate values (maximum likelihood or multiple imputation).
\item \textit{How to impute missing values?} We compare different single imputation/matrix completion methods (for instance using conditional models, low-rank models, etc.).
\item \textit{How to predict with missing values?} We consider building predictive models, e.g. using random forests \citep{breiman_2001}, on data with incomplete predictors. The workflows present different strategies to deal with missing values in the covariates both in the training set and in the test set.
\end{itemize}

The aim of these workflows is threefold: 1) they provide a practical
implementation of concepts and methods discussed in the lectures and
bibliography sections of the platform; 2) they are implemented in a generic way,
allowing for re-use on other datasets, for integration of other
estimation or imputation methods; 3) the distinction between inference, imputation, and prediction lets the user keep in mind the solutions are not the same.

Furthermore, the workflows allow for a transparent and open discussion about the proposed implementations, which can be followed on the project GitHub repository, referencing proposals and discussions about practicable extensions of the workflows.

Additionally, a workflow on \textit{How to do causal inference with incomplete covariates/attributes in R?} demonstrates simple weighting and doubly robust estimators for treatment effect estimation using R. This workflow is based on the R implementation of the methodology proposed by \citet{Mayer2020}.

We provide a more detailed view on the proposed workflows in a later section, with examples of tabular or graphical outputs that can be obtained as well as recommendations on how to interpret and leverage these outputs.

\subsection{Missing values lectures}\label{sec:lectures}

For someone unfamiliar with missing data, it is a challenge to know where to begin the journey of understanding them, and the methods to handle them.
This challenge is addressed with \samp{R-miss-tastic}, which makes the material to get started easily accessible.

Teaching and workshop material takes many forms -- from slides, course notes, lab workshops, video tutorials and in-depth seminars. The material is of high quality, and has been generously contributed by numerous renowned researchers who investigate the problems of missing values, many of whom are professors having designed introductory and advanced classes for statistical analyses with missing data. This makes the material on the \samp{R-miss-tastic} platform well suited for both beginners and more experienced users.

These teaching and workshop materials are described as \samp{lectures}, and are organized into five sections:

\begin{enumerate}
\item General lectures: introduction to statistical analyses with missing values; the role of visualization and exploratory data analysis for understanding missingness and guiding its handling; theory and concepts are covered, such as missing values mechanisms, likelihood methods, and  imputation.
\item Multiple imputation: introduction to popular methods of multiple imputation (joint modeling and fully conditional), how to correctly perform multiple imputation and limits of imputation methods.
\item Principal component methods:
introduction to methods exploiting low-rank type structures in the data for
visualization, imputation and estimation.
\item Specific data or applications types: lectures covering in details various sub-problems such as missing values in \textit{time series}, in \textit{surveys}, or in treatment effect estimation (\textit{causal inference}). Indeed, certain data types require adaptations of standard missing values methods (for instance handling time dependence in time series \citep{moritz_2017}) or additional assumptions about the impact of missing values (such as the impact on confounded treatment effects in causal inference  \citep{Mayer2020}). But also more in-depth material, for instance video recordings from a virtual workshop on \textit{Missing Data Challenges in Computation, Statistics and Applications}\footnote{\url{https://www.ias.edu/math/mdccsa}} held in 2020.
\item Implementations: a non-exhaustive list of detailed vignettes 
describing functionalities of {R} packages and of {Python} modules that implement some of the statistical 
analysis methods covered in the other lectures. For instance, the functionalities and possible applications of the \pkg{missMDA} {R} package are presented in a brief summary, allowing the reader to compare the main differences between this package and the \pkg{mice} package which is also summarized using the same summary format.
\end{enumerate}

\begin{figure}
\begin{center}
\begin{subfigure}[b]{0.63\textwidth}
\includegraphics[width=\textwidth]{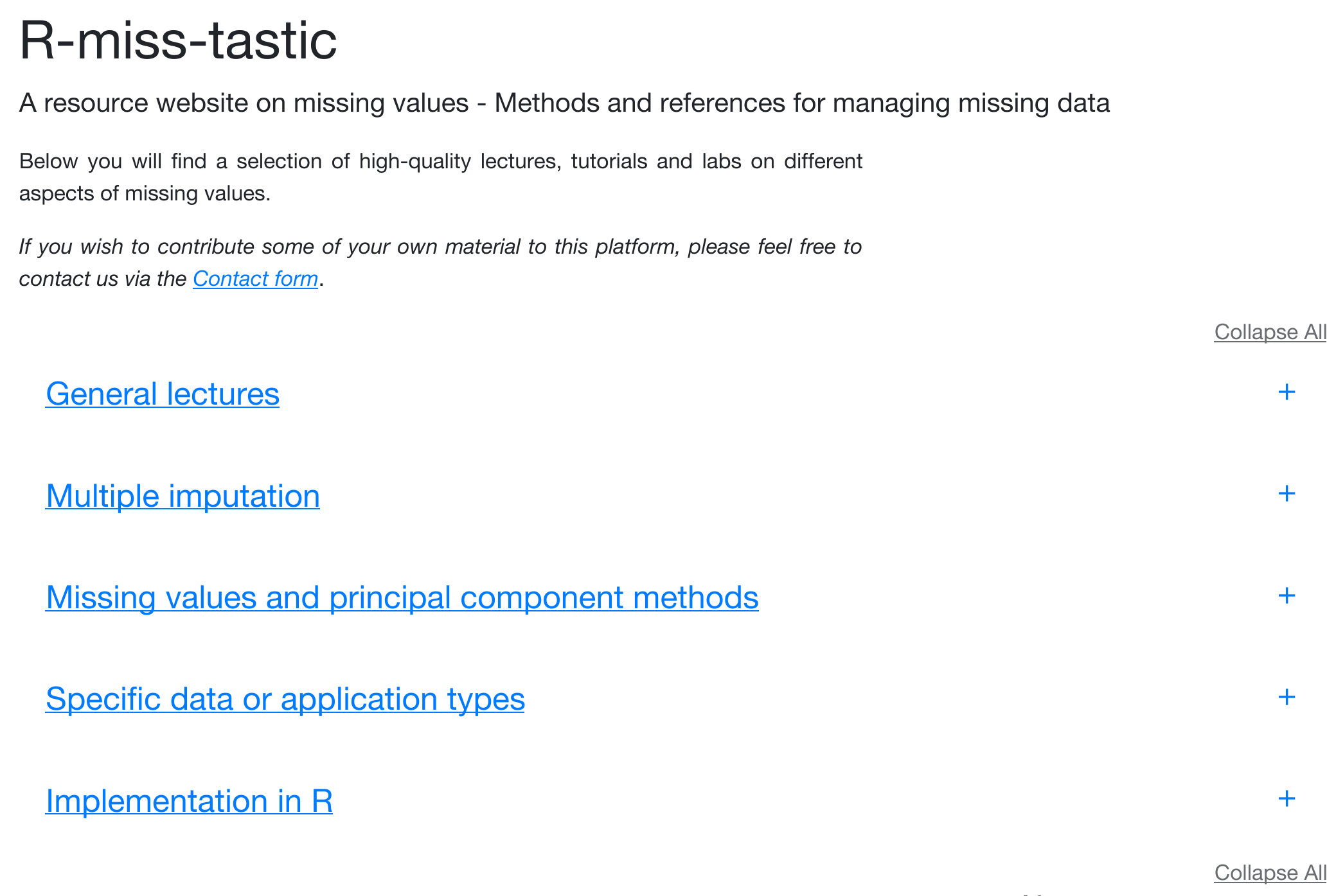}
\caption{Collapsed view}
\end{subfigure}
\begin{subfigure}[b]{0.36\textwidth}
\includegraphics[width=\textwidth]{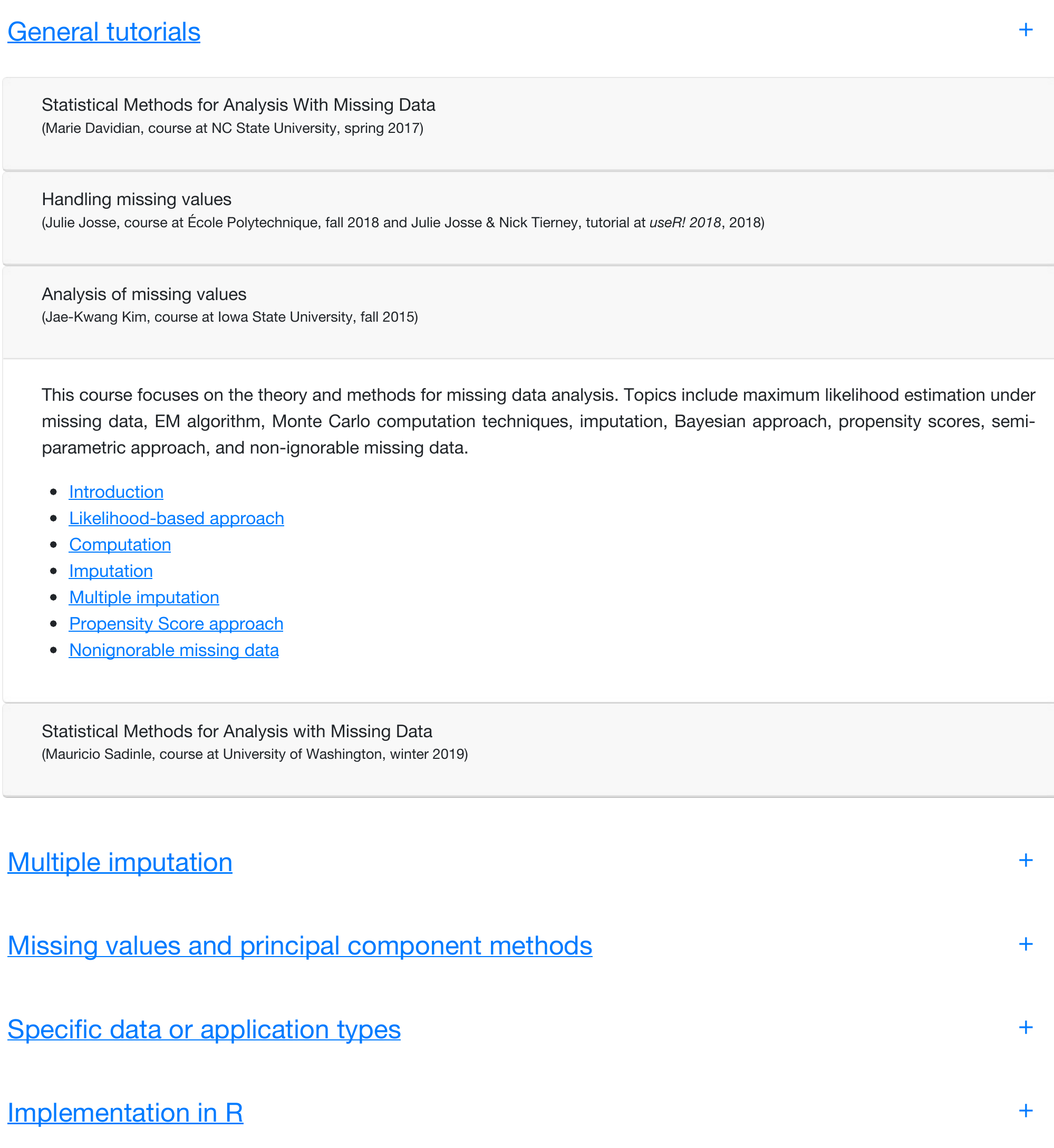}
\caption{Extract}
\end{subfigure}
\end{center}
\caption{Lecture overview. \label{fig:lectures}}
\end{figure}

Figure~\ref{fig:lectures} illustrates two views of the lectures page: Figure~\ref{fig:lectures}A shows a collapsed view presenting the different topics, Figure~\ref{fig:lectures}B shows an example of the expanded view of one topic (General tutorials), with a
detailed description of one of the lecture (obtained by clicking on its title),
\samp{Analysis of missing values} by Jae-Kwang Kim. Each lecture can contain
several documents (as is the case for this one) and is briefly described by
a header presenting its purpose.

Lectures that we found very complete and thus highly recommend are:
\begin{itemize}
\item \textit{Statistical Methods for Analysis with Missing Data} by Mauricio
Sadinle (in \samp{General tutorials});
\item \textit{Missing Values in Clinical Research -- Multiple Imputation} by
Nicole Erler (in \samp{Multiple imputation});
\item \textit{Handling missing values in PCA and MCA} by Fran\c cois Husson. (in
\samp{Missing values and principal component methods});
\item \textit{Modern use of Shared Parameter Models for Dropout (in longitudinal and time-to-event data)} by Dimitris Rizopoulos (in \samp{Specific data or application types}).
\end{itemize}

The purpose of these lectures is to provide either an introduction or a deeper understanding of the statistical problems and proposed solutions in terms of their (mathematical) derivation and theoretical scope. So there is less focus on practical demonstrations with real data, or a systematic comparison of all methods for the same problems. This is covered in the section presenting in detail the developed workflows.

\subsection{References on missing values}\label{sec:biblio}

\begin{figure}
\begin{center}
\begin{subfigure}[b]{0.515\textwidth}
\includegraphics[width=\textwidth]{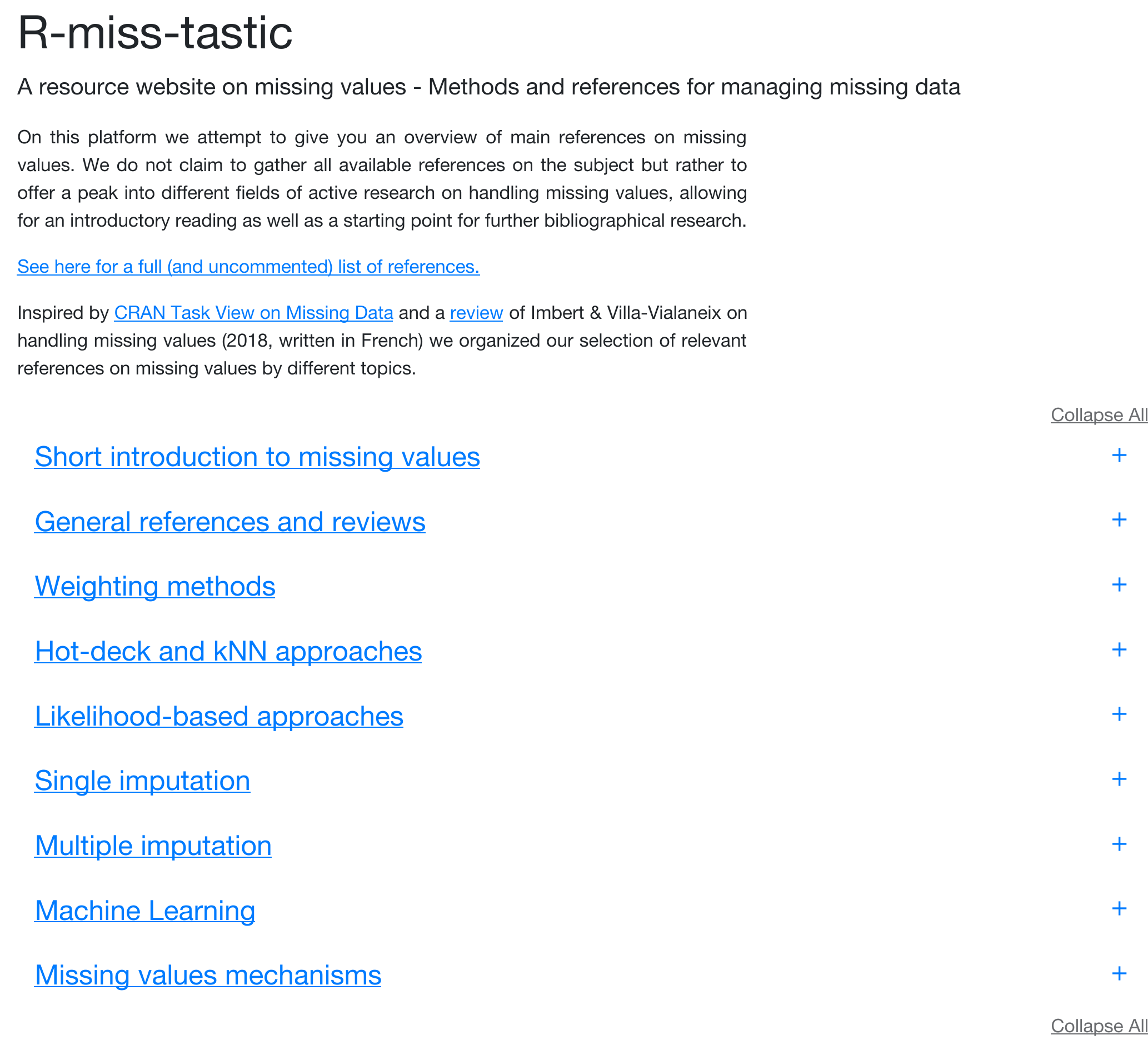}
\caption{Contextualized version}
\end{subfigure}
\begin{subfigure}[b]{0.475\textwidth}
\includegraphics[width=\textwidth]{local-bibliography-1.pdf}
\caption{Unordered version}
\end{subfigure}
\end{center}
\caption{Bibliography overview. \label{fig:bibliography}}
\end{figure}

Complementary to the \textit{Lectures} section, this part of the platform serves as a
broad overview on the scientific literature discussing missing values taxonomies
and mechanisms and statistical, machine learning methods to handle them. This overview covers both
classical references to books, articles, etc. such as \citet{Schafer2002,VanBuuren2012, Carpenter2012,little2019statistical} and more recent developments such as
\citet{josse_etal_2019, gondara_wang_2018}, which study the consistency of supervised learning with missing values. The entire (non-exhaustive)
bibliography can be browsed in two ways: 1) a complete list,
filtered by publication type and year, with a search option for the authors or, 2) as a contextualized version. For 2), we classified the references into
several domains of research or application, briefly discussing important
aspects of each domain. This dual representation is shown in
Figure~\ref{fig:bibliography} and allows for an extensive search in the existing
literature, while providing some guidance for those focused on a specific
topic. All references are also collected in a unique BibTex file made available
in the GitHub repository\footnote{in \url{resources/rmisstastic_biblio.bib}}. This shared file allows external users to easily propose additions to the bibliography, which are then reviewed by the platform committee, composed of researchers with different focuses on  missing values.

\subsection{Missing values implementations}

\paragraph*{{R} packages}

As mentioned in the introduction, the platform development is based on the release of the \ctv{MissingData} CRAN Task View, which
currently lists approximately 150 {R} packages. The CRAN Task View is continuously updated, adding new {R} packages, and removing obsolete ones. Packages are organized by topic: \textit{exploration of missing data},
\textit{likelihood based approaches},
\textit{single imputation}, \textit{multiple imputation}, \textit{weighting methods},
\textit{specific types of data}, and \textit{specific application fields}.
We selected only sufficiently mature and stable packages already published on CRAN or Bioconductor. This ensures  all listed
packages can easily be installed and used by practitioners.

\begin{figure}
\begin{center}
\begin{subfigure}[b]{0.575\textwidth}
\includegraphics[width=0.9\textwidth]{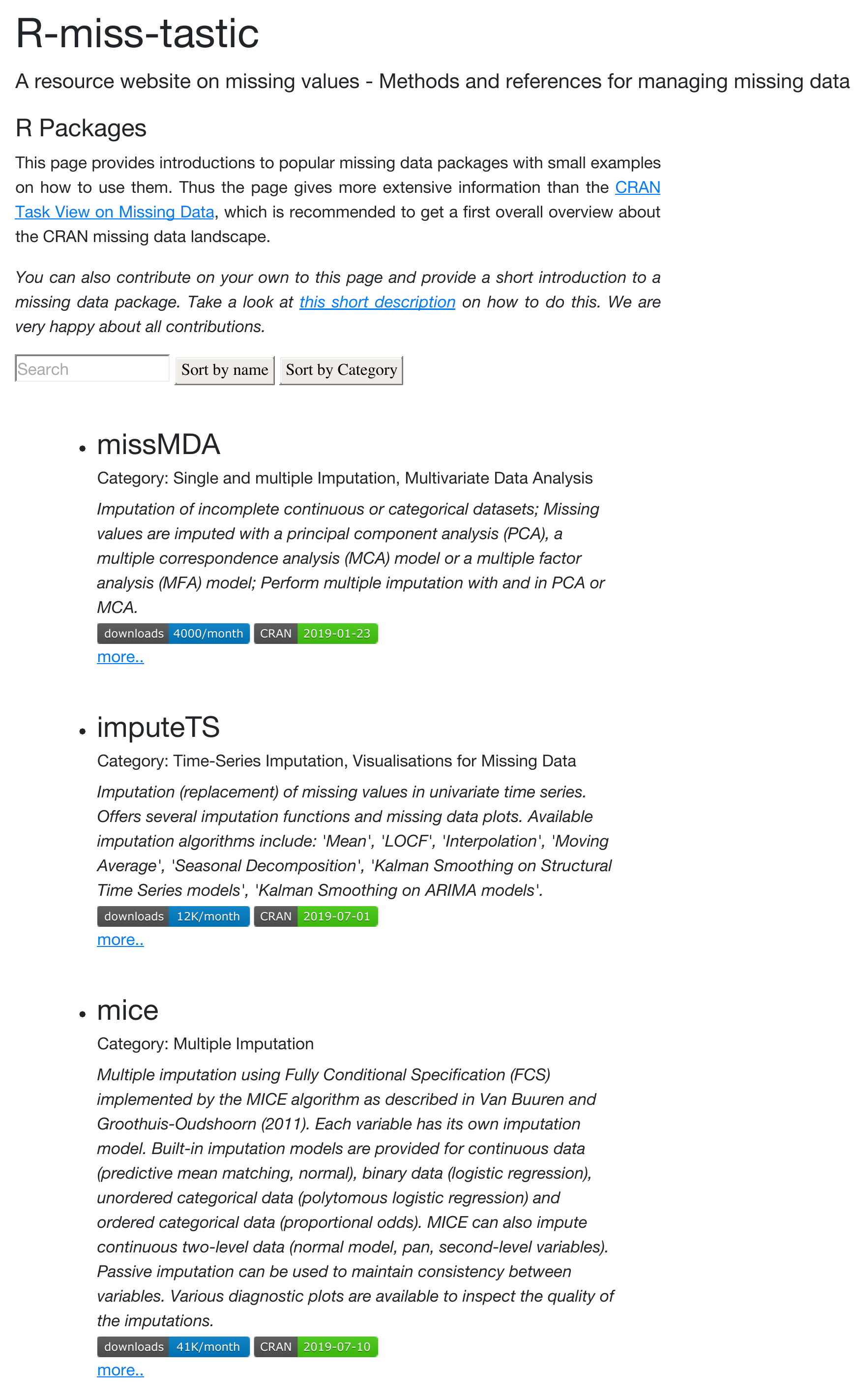}
\caption{Extract of global view}
\end{subfigure}
\begin{subfigure}[b]{0.415\textwidth}
\includegraphics[width=0.9\textwidth]{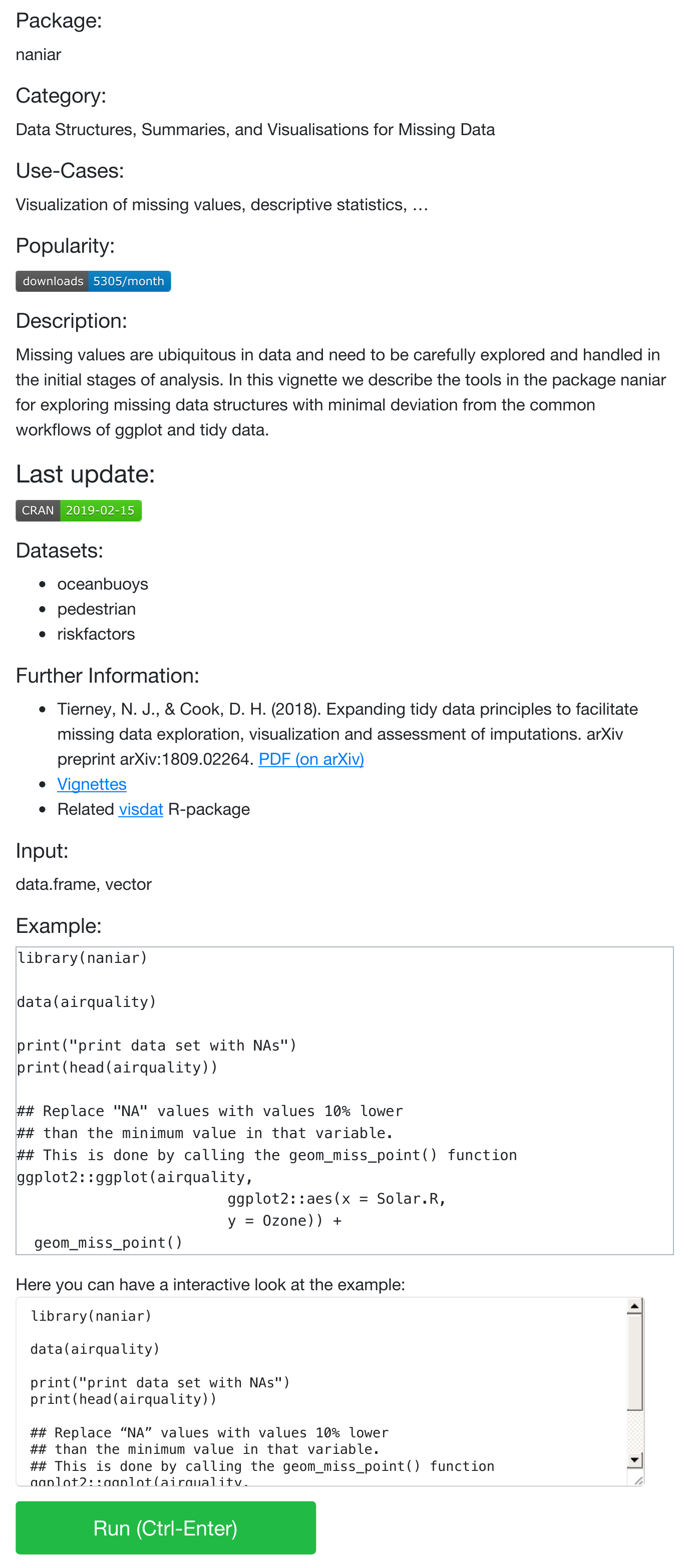}
\caption{Description sheet}
\end{subfigure}
\end{center}
\caption{{R} packages overview. \label{fig:packages}}
\end{figure}

Even though the Task View classifies packages into different sub-domains, it may still be a challenge for practitioners and researchers inexperienced with missing values to choose the most relevant package for their application. To address this challenge, we provide a partial and slightly more detailed 
overview of existing {R} packages, selecting the most popular and versatile ones. This overview is a blend of the
Task View, and of the individual package description pages and vignettes as provided on CRAN or Bioconductor. For each
selected package (7 at the date of writing of this article: \pkg{imputeTS}, \pkg{mice}, \pkg{missForest}, \pkg{missMDA}, \pkg{naniar}, \pkg{simputation} and \pkg{VIM}), we provided a category (in the style of the categorization in the Task View), a
short description of use-cases, its description (as on CRAN), the usual CRAN
statistics (number of monthly downloads, last update), the
handled data formats (e.g., \texttt{data.frame}, \texttt{matrix},
\texttt{vector}), a list of implemented algorithms (e.g., k-means,
PCA, decision tree), the list of available datasets, some references (such as articles and books), and a small chunk of code, ready for a direct execution on the platform
via the \textit{{R} package Documentation}\footnote{https://rdrr.io/snippets/}.
Figure~\ref{fig:packages} shows the condensed view of the package page and the
expanded description sheet of a given package (here \pkg{naniar}).

We believe shortlisting {R} packages is highly useful for
practitioners new to the field, as it demonstrates data analysis that handles  
missing values in the data. We are aware that this selection is subjective, and we
welcome external suggestions for other packages to add to this shortlist.

\paragraph*{{Python} modules}

To the best of our knowledge, very few methods are already implemented for handling missing data in {Python}. However, one of the major libraries for machine learning and data analysis, \texttt{scikit-learn} \citep{pedregosa2011scikit} has recently proposed a module for simple and multiple imputations, \texttt{sklearn.impute}. Also, as an alternative, the \texttt{statsmodels}\footnote{\url{https://www.statsmodels.org/stable/about.html}} library also has an implementation module for multiple imputation in {Python} now. Additionally, the \texttt{missingno} toolset \citep{Bilogur2018} facilitates visualizing missing values missing values for exploratory data analyses. We regularly survey new {Python} implementations for handling missing values and, if pertinent, from a theoretical and practical point of view, reference them on our platform. We expect this to promote their use but also additional assessment by practitioners and researchers from the missing values (statistics/machine learning) community.

\subsection{Datasets}

\begin{figure}
\begin{center}
\includegraphics[width=0.6\textwidth]{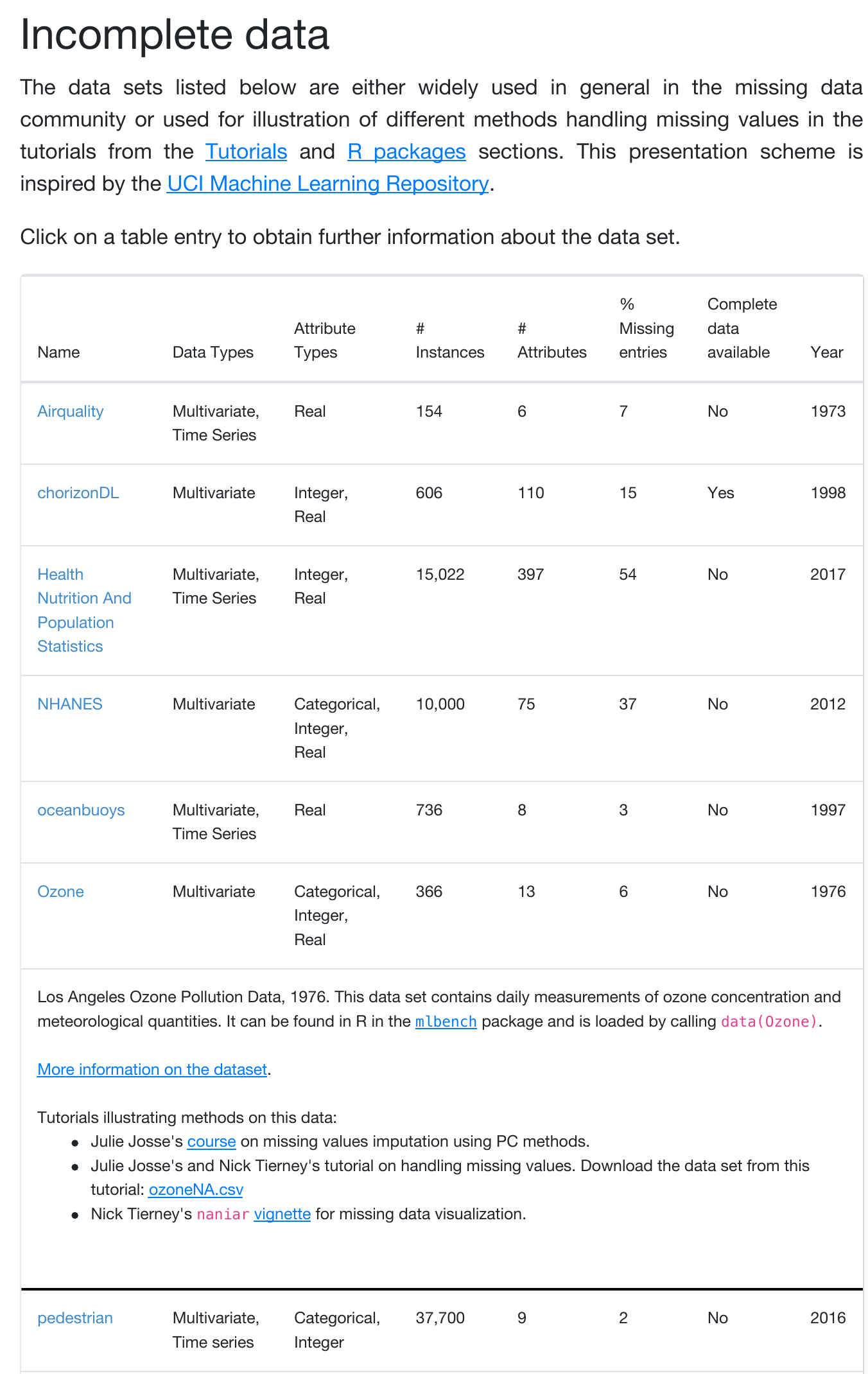}
\end{center}
\caption{Datasets overview. \label{fig:datasets}}
\end{figure}

Especially in methodology research, an important aspect is the comparison of
different methods to assess their respective strengths and weaknesses. Several datasets
are recurrent in the missing values literature but have not been referenced together yet. We gathered publicly
available datasets that have recurrently been used for comparison or
illustration purposes in publications, {R} packages and tutorials. Most of these
datasets are already included in {R} packages but some are available in other
data collections. Figure~\ref{fig:datasets} shows how the datasets are
presented, with a detailed description shown for one of the dataset (\samp{Ozone},
obtained by clicking on its name). The description follows the UCI Machine
Learning Repository presentation \citep{dua_2019}, including a short
description of the dataset, how to obtain it, external references describing
the dataset in more details, and links to tutorials/lectures on our websites or
to vignettes in {R} packages that use the dataset.

In addition, the \textit{Datasets} section also references existing methods for generating missing data, given assumptions on their generation
mechanisms (as in the {R} package \pkg{mice}). 

Note, however, that the list of datasets gathered here is short compared to benchmark datasets for full data methods such as the UCI Machine Learning Repository. Therefore, our proposed list also serves as an invitation to tackle this lack of a wider variety of common benchmark datasets in the missing data community. 

\subsection{Additional content}

This unified platform collects and edits the contributions of numerous individuals who have investigated missing values problems, and developed methods to handle them. To
provide an overview of some of the main actors in this field, the list of all
contributors who agreed to appear on this platform is given with links to their
personal or to their research lab website.

We also provide links to other interesting websites or working groups, not 
necessarily working with {R} and {Python} \citep{python} but with other programming languages such as 
SAS/STAT$^{\textrm{\tiny\textregistered}}$ and STATA \citep{stata}.

Two other features are finally provided to engage the community:
\begin{enumerate}
  \item a regularly updated list of events such as conferences or summer 
schools with special focus on missing values problems, and
  \item a list of recurring questions together with short answers and links for 
more details for every question (FAQ).
\end{enumerate}

\section{Details of missing values workflows} \label{sec:workflows}

After this general introduction to the \samp{R-miss-tastic} project and platform and the overview of its structure, we now turn to a more detailed presentation of the various workflows that we have developed and proposed on this platform. 

To allow for both hands-on tutorials illustrating current practices and state-of-the-art and ready-to-use pipelines, we propose the workflows under different formats such as HTML, PDF, {R} Markdown (for {R} code) and IPython Notebook (for {Python} code). 
We encourage practitioners and researchers to
use and adapt these workflows and propose modifications and improvements, in order to increase reproducibility and comparability of their work. Of course, we are aware that these workflows do not cover the entire spectrum of existing methods 
and data problems. The goal of the proposed workflows is rather to initiate a joint effort to create a larger spectrum of open-source workflows, and to encourage the use of standardized procedures to handle missing values. 
With an incomplete dataset at hand, prior to embarking on an in-depth statistical analysis, two preliminary steps are essential: (i) a descriptive analysis leveraging visualization packages such as \pkg{VIM} \citep{VIM} or \pkg{naniar} \citep{naniar}; (ii) a specific aim has to be defined in order to choose a specific method to use. 

An example of a method whose success crucially depends on the analyst's goal is \textit{mean imputation}: this approach is strongly counter-indicated if the aim is to estimate parameters, but it can be consistent if the aim is to predict as well as possible \citep{josse_etal_2019}. 
Following this observation, our workflows are divided into different parts, defined by the objective of the statistical analysis. We aim to present and compare the main implementations available both in {R} and {Python} for each objective. Currently there are seven workflows available on the platform and we briefly present their scope and use below. For details on the implementations we encourage the reader to open the corresponding workflows, all available on the \samp{R-miss-tastic} platform.


\subsection{How to generate missing values?}
\label{sec:howtogen}

The goal of these workflows is to propose functions to generate missing values under different mechanisms. This code aims to unify classical solutions to do this. Indeed, a usual strategy to compare imputation or estimation strategies is to introduce (additional) missing values in the dataset, and use the ground truth for these missing values to evaluate the strategies (see the following section).

\citet{rubin_B1976} classifies the cause for a lack of data into three missing data mechanisms. The missing data mechanism is said to be: (i) missing completely at random (MCAR) if the lack of data is totally independent of the data values, (ii) missing at random (MAR) if the process that causes the missing data only depends on the observed values and (iii) missing not at random (MNAR) if the unavailability of the data depends on the missing variables. See \citet{sportisse2021handling} for a recent overview on the topic. 
\paragraph*{In {R}} In the {R} \href{https://rmisstastic.netlify.app/how-to/generate/misssimul}{workflow}\footnote{\url{https://rmisstastic.netlify.app/how-to/generate/misssimul}}, we have implemented the main function \texttt{produce\_NA}\footnote{\url{https://rmisstastic.netlify.app/how-to/generate/amputation.R}} which facilitates generating missing values under the three missing data mechanisms outlined above. This function internally calls the \texttt{ampute} function from the \pkg{mice} package \citep{mice} but we chose to simplify its use while adding some additional options to specify the missing values generation. In addition, the original \texttt{ampute} function generates missing values only for a complete dataset with quantitative variables\footnote{If qualitative variables have previously been transformed by one-hot-encoding, they can also be handled by the \texttt{ampute} function of \pkg{mice}. The \texttt{produce\_NA} function internally handles the transformation of qualitative variables prior to amputation.}. In the main function of our workflow, the user can easily introduce (additional) missing values in a complete or incomplete dataset composed of quantitative, categorical, or mixed variables, by choosing the mechanism and the percentage of missing values to be introduced. The function then returns the data matrix containing the new dataset with missing values,
that also includes the missing values already present in the input data, and the indicator matrix (a binary matrix where an entry is equal to $1$ if a new missing value has been generated at the same location in the data matrix and $0$ otherwise). 

The three main arguments are the initial dataset (\texttt{data}) in which the missing values are introduced using a given missing data mechanism (\texttt{mechanism}) and a given percentage of missing values (\texttt{perc.missing}). For example, to introduce 20\% of MCAR values in the dataset $X$, the code is detailed below. 
\begin{example}
X.miss.mcar <- produce_NA(data = X, mechanism = "MCAR", perc.missing = 0.2)
X.mcar <- X.miss.mcar['data.incomp']
R.mcar <- X.miss.mcar['idx_newNA']
\end{example}

For instance, if $X$ contains three variables (fully observed) denoted as $X_{1},\ X_{2},\ X_{3}$, two options are available to generate MAR values:
\begin{enumerate}
    \item The first option consists of generating missing values in $X_{1}$ by using a logistic model depending on $(X_2,X_3)$, which are fully observed, i.e., 
    \begin{equation}\label{eq:logmod}
    \mathbb{P}(R_1=0|X;\phi)=1/(1+\exp(-(\phi_2 X_2+\phi_3 X_3)),
    \end{equation}
    where $\phi=(\phi_2,\phi_3)$ is the parameter of the missing data mechanism. In our function, $\phi$ is chosen such that the given percentage of missing values is achieved. This allows us to obtain missing values in the first variable $X_1^{\mathrm{NA}}$. Then, the same strategy is performed to introduce missing values in $X_2$ and $X_3$, by using a logistic model depending on $(X_1,X_3)$ (fully observed) and $(X_1,X_2)$ (fully observed) respectively. To get the final matrix containing missing values, we concatenate $X_1^{\mathrm{NA}}$, $X_2^{\mathrm{NA}}$ and $X_3^{\mathrm{NA}}$ by handling the rows containing only missing values. 

    \item The second option consists of generating the missing values \textit{by pattern}, i.e., by rows. In this case, the combinations of which variables are observed and missing are specified in a pattern matrix. For the MAR mechanism, in each pattern, at least one variable must be observed. 
    An example (the choice by default) of such a pattern matrix is 
    $$\begin{pmatrix}
    0 & 1 & 1 \\
    1 & 0 & 1 \\
    1 & 1 & 0 \\
    \end{pmatrix},$$ where $0$ indicates that the variable should have missing values whereas $1$ means that it should be observed. For example, the first pattern means that the process which causes the missingness of the first variable $X_1$ depends on the values of $X_2$ and $X_3$ which are observed. 
\end{enumerate}

We also propose several ways to generate missing values, under the MNAR mechanism. It includes the most general one when the missingness depends on both the missing variables and the observed variables. It also includes the self-masked mechanism, where the unavailability of the data only depends on their values themselves.
For example, it is possible to introduce self-masked missing values using a quantile censorship for which the form is precised by the argument \code{self.mask}, e.g., if set to \code{\samp{lower}}, then the values are amputed based on a quantile from the lower tail of the empirical distribution such that the target proportion of missing values is achieved.

\paragraph*{In {Python}} To our knowledge, there is no specific module in {Python} to generate missing values.
Consequently, we implemented such functions, in a {Python} workflow, which similarly to its {R} counterpart  \href{https://rmisstastic.netlify.app/how-to/python/generate_html/how\%20to\%20generate\%20missing\%20values}{workflow}\footnote{\url{https://rmisstastic.netlify.app/how-to/python/generate_html/how\%20to\%20generate\%20missing\%20values}} allows us to generate missing values under by different mechanisms and different percentage of missing values.\footnote{The code has been partially developed in collaboration with Boris Muzellec (Inria Paris).} The key difference with the {R} workflow is that the data set must be complete and can currently only contain quantitative variables. For MAR and MNAR mechanisms, only the option \textit{not by pattern} has been implemented. In this case, for a dataset $X$ with three variables, a variable is chosen to be fully observed (say $X_3$), and the process which causes the missingness of two other variables ($X_1$ and $X_2$) depends on the values of the fully observed variable, for example with the logistic model given in \eqref{eq:logmod}.

\subsection{How to impute missing values?}
\label{sec:howtoimp}

The aim of these workflows (in {R} and {Python}) is to compare the most classical imputation methods and to propose a reference pipeline for comparison on simulated and real datasets, which can be easily extended with other imputation methods. 
Here, the imputation methods are considered as such, i.e., the objective is not to estimate a parameter or to perform a statistical analysis on a completed dataset but to impute missing values to get a complete dataset in the best possible way. Therefore, we evaluate the methods in terms of imputation quality, by using the mean squared error (MSE). More precisely, the procedure is the following one: (i) we have access to a complete dataset $X$, (ii) missing values are introduced in $X$ and we get an incomplete dataset $X^{\mathrm{NA}}$, (iii) this incomplete dataset is imputed and we obtain an imputed dataset $X^{\mathrm{imp}}$, (iv) the MSE, which measures the error committed by the imputation of the missing values, is computed: it is the $\ell_2$-norm of the difference of the imputed dataset and the complete one). Note that this procedure can also be performed on an incomplete dataset by introducing additional missing values. However, for now, both {R} and {Python} workflows only consider complete datasets.

Different types of imputation methods are included in this workflow:
\begin{enumerate}
    \item \underline{imputation by the mean}, which serves as a naive baseline. 
    \item \underline{conditional models}, if the imputation relies on the conditional expectation or a draw from the conditional distribution of each variable given the others.
    \begin{itemize}
        \item in {R}: 
        \begin{itemize}
            \item \pkg{mice} \citep{mice}: a multiple imputations method by chained equations.  Even if it is a returns several imputed data sets, they can be aggregated using the mean of the imputations to get a single imputation. 
            \item \pkg{missForest} \citep{stekhoven2012missforest}: imputes iteratively by training random forests.
        \end{itemize}
        \item in {Python}:
        \begin{itemize}
            \item \texttt{IterativeImputer} of scikit-learn library \citep{pedregosa2011scikit}: this function is inspired by mice, but it uses (iterative) regularized regression, imputing by the conditional expectation, and providing a simple imputation. We also use the \texttt{ExtraTreesRegressor} estimator of \texttt{IterativeImputer}, which trains iterative random forests (it is similar to \pkg{missForest} in {R}).
        \end{itemize}
    \end{itemize}
    \item \underline{low-rank based models}, the data matrix to impute is assumed to be generated as a low rank structure plus a noise term.
    \begin{itemize}
        \item in {R}:  
        \begin{itemize}
            \item \pkg{softImpute} \citep{hastie2015matrix}: minimizes the re-weighted least squares error penalized by the nuclear norm.
            \item \pkg{missMDA} \citep{josse2016missmda}: minimizes the re-weighted least squares error penalized by a mix between the $\ell_2$-norm and $\ell_0$-norm. 
        \end{itemize}
        \item in {Python}: \pkg{softImpute} (coded for the purpose of this notebook and available  \href{https://github.com/R-miss-tastic/website/blob/master/static/how-to/python/softimpute.py}{here}\footnote{\url{https://github.com/R-miss-tastic/website/blob/master/static/how-to/python/softimpute.py}}), which minimizes the re-weighted least squares error penalized by the nuclear norm and with an internal cross-validation step to choose the regularization parameter. 
    \end{itemize}
    \item \underline{machine learning methods} (for the {Python} workflow only) using optimal transport or variational autoencoders.
    \begin{itemize}
        \item in {Python}:  
        \begin{itemize}
            \item \texttt{MIWAE} \citep{mattei2019miwae}: imputes missing values with a deep latent variable model based on importance weighted variational inference.
            \item \texttt{Sinkhorn} \citep{muzellec2020missing}: randomly extracts several batches and minimizes optimal transport distances between batches to impute missing values. 
        \end{itemize}
    \end{itemize}
\end{enumerate}
For the sake of clarity, we show a comparison table (Table \ref{tab:compar}) in the Appendices, showing the difference of scope between R and Python packages used in the R-miss-tastic workflows.

\paragraph*{In {R}} This \href{https://rmisstastic.netlify.app/how-to/impute/missimp}{workflow}\footnote{\url{https://rmisstastic.netlify.app/how-to/impute/missimp}} provides two main functions which compares the imputation methods: (i) on a simulated dataset for different mechanisms and percentage of missing values (\texttt{how\_to\_impute}) or (ii) on a list of real datasets and a given mechanism and percentage of missing values (\texttt{how\_to\_impute\_real}). 

\begin{figure}
\begin{subfigure}[b]{1\textwidth}
\centering
\includegraphics[width=0.8\textwidth]{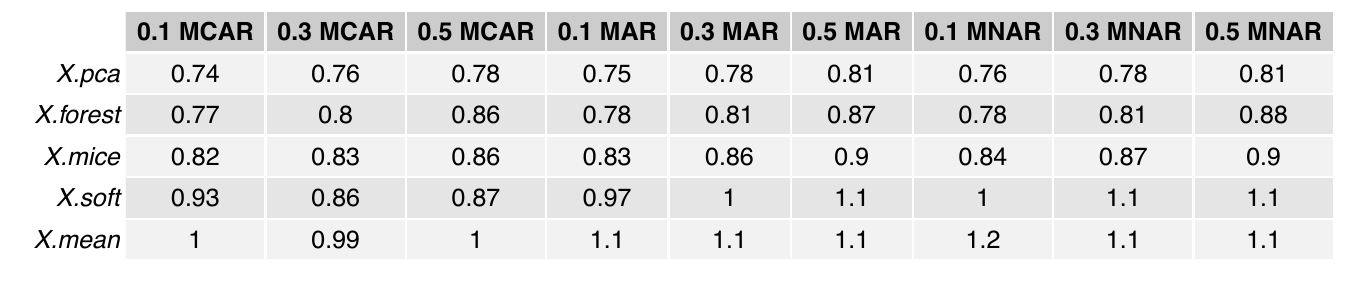}
\caption{\label{fig:howtoimpute_R_table} Output of the function \texttt{how\_to\_impute} in {R}. The results for the MSE are truncated to two digits. Note that the line \code{X.pca} is the result for \pkg{missMDA}. For all methods, the default parameter choices are used.}
\end{subfigure}
\begin{subfigure}[b]{1\textwidth}
\centering
\includegraphics[width=0.8\textwidth]{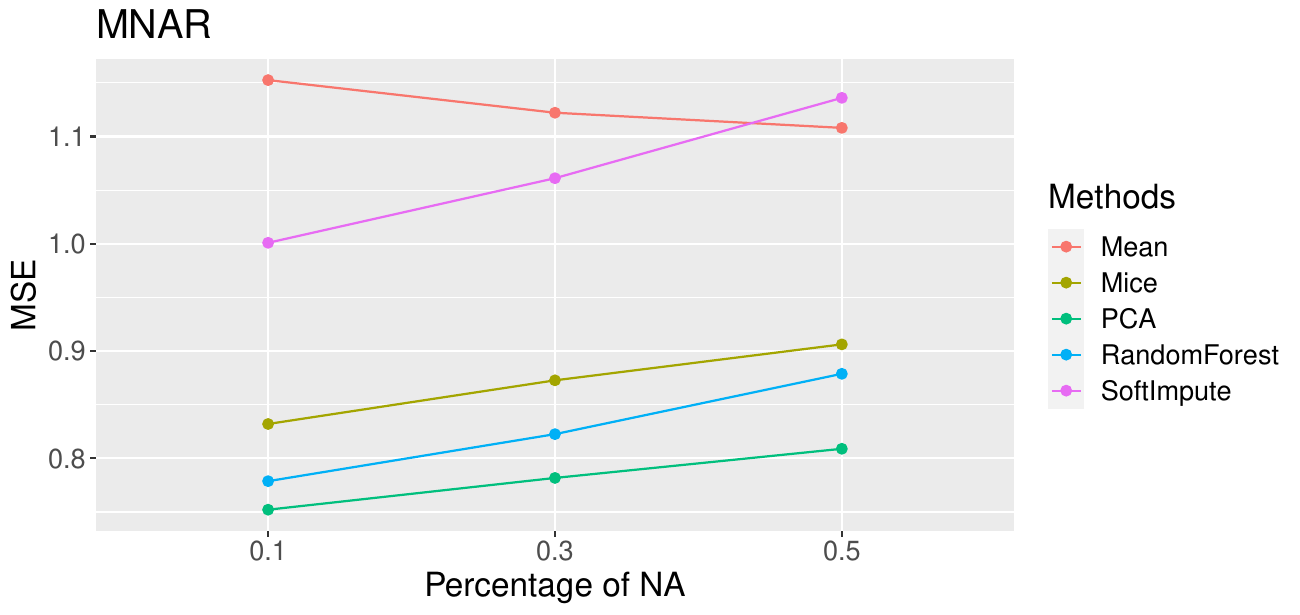}
\caption{\label{fig:howtoimpute_R_fig} Example of plot for the MNAR mechanism.}
\end{subfigure}
\caption{\label{fig:howtoimpute_R_main}Tabular and graphical outputs of the {R} function \texttt{how\_to\_impute}. The methods \pkg{mice}, \pkg{missForest}, \pkg{softImpute} and \pkg{missMDA} are compared with the naive imputation by the mean for several percentages of missing values (10\%, 30\%, 50\%). The mean of the MSEs computed for several generations of missing values are given. In the tabular, the results are shown for several mechanisms (MCAR, MAR, MNAR) and the plot corresponds to the MNAR mechanism.}
\end{figure}

The function \texttt{how\_to\_impute} takes as input a complete dataset (\texttt{X}), a list of percentages of missing values (\texttt{perc.list}) and a list of missing data mechanisms (\texttt{mecha.list}). The code to use this function is given below.
\begin{example}
perc.list <- c(0.1, 0.3, 0.5) 
mecha.list <- c("MCAR", "MAR", "MNAR") 
res <- how_to_impute(X = X, perc.list = perc.list, mecha.list = mecha.list, nbsim = 10) 
\end{example}
The output of the first function \texttt{how\_to\_impute} is the mean of the methods' MSEs for the different missing values settings by taking the average over several repetitions (the number of repetitions can be specified through the argument \code{nbsim}). Figure \ref{fig:howtoimpute_R_main} shows the output of this function and its associated plot, when the simulated dataset is Gaussian with $n=1000$ observations, $d=10$ variables, a mean vector such that $\mu_i=1, \: \forall i \in \{1,\dots,d\}$ and a covariance matrix such that $\Sigma_{ij}=0.5 \textrm{ if } i\neq j \in \{1,\dots,d\}$, and $\Sigma_{ij}=1 \textrm{ if } i=j$. First, the mean of the methods' MSEs for the different missing values settings are reported in Figure \ref{fig:howtoimpute_R_table}. We can remark that for the MCAR mechanism, the methods perform well, while for the MNAR mechanism, the results are generally closer to those of the naive imputation by the mean. As expected, most methods give worse results for high percentages of missing values. Besides, Figure \ref{fig:howtoimpute_R_fig} shows one of the associated plot for MCAR data (there is also a plot for MAR data and a plot for MNAR data). In the first part of the appendix, this function is illustrated for a particular dataset and the code to obtain Figure \ref{fig:howtoimpute_R_main} is given.

\begin{figure}
    \centering
    \includegraphics[width=0.6\textwidth]{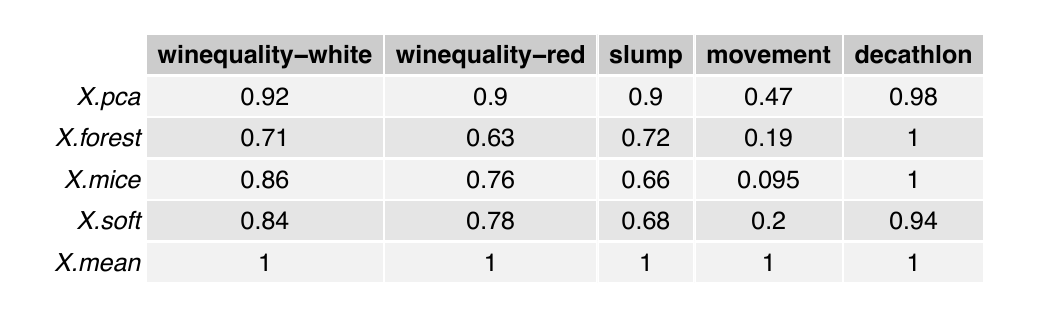}
    \caption{\label{fig:howtoimpute_real} Output of the {R} function \texttt{how\_to\_impute\_real}. The results for the MSE are truncated to two digits. The methods \pkg{mice}, \pkg{missForest}, \pkg{softImpute} and \pkg{missMDA} for several real datasets in which 20\% MCAR missing values have been introduced.}
\end{figure}

The second function \texttt{how\_to\_impute\_real} takes as input a list of datasets (\texttt{datasets\_list}), a list of missing data mechanisms (\texttt{mech}) and a given percentage of missing values (\texttt{perc}). It returns a table containing the mean of the MSEs for the simulations performed and a table for the summary plot shown in Figure \ref{fig:howtoimpute_real}. This can be particularly useful for practitioners who would like to have an indication of which method might be the most suited for a given or for several specific datasets.
Here, the real datasets are taken from the \href{http://archive.ics.uci.edu/ml/index.php}{UCI repository}\footnote{\url{http://archive.ics.uci.edu/ml/index.php}} \citep{dua_2019}. An example of how to use this function in practice is detailed below.

\begin{example}
datasets_list <- list(wine_white = wine_white, wine_red = wine_red, slump = slump, 
                        movement = movement, decathlon = decathlon)
names_dataset <- c("winequality-white", "winequality-red", "slump", "movement", "decathlon") 
perc <- 0.2
mecha <- "MCAR" 
res <-  how_to_impute_real(datasets_list = datasets_list ,perc = perc, mech = mecha, 
                                    nbsim = 10, names_dataset = names_dataset)
\end{example}

An additional \href{https://rmisstastic.netlify.app/how-to/external/comparison_imputation_deep_classical}{workﬂow}\footnote{This workflow has been implemented by an external contributor, Fran\c{c}ois Husson (Professor in Statistics, France).} is available and compare other deep-learning imputation strategies to most classical ones on data sets simulated either with linear relationships and nonlinear relationships. The conclusions points to better behavior of the low-rank based imputations methods even when deep-learning methods are tuned.

\paragraph*{In {Python}} The {Python} \href{https://rmisstastic.netlify.app/how-to/python/howtoimpute}{workflow} is very similar to its R counterpart. The two same functions, \texttt{how\_to\_impute} and \texttt{how\_to\_impute\_real}, have been implemented.

\subsection{How to estimate parameters with missing values in {R}?}

This R \href{https://rmisstastic.netlify.app/how-to/estimate/missestim}{workflow}\footnote{\url{https://rmisstastic.netlify.app/how-to/estimate/missestim}} is dedicated to a specific inferential framework when the aim is to estimate linear and logistic regression parameters for multivariate normal data. It is currently only available in {R}, as there are no analogous implementations available in {Python} to our knowledge. 

There are two main methods to estimate parameters with missing values: maximum likelihood estimation adapted to missing values, using, e.g., EM-based algorithms or using multiple imputation.
In this workflow, we compare two instances of these main methods, using available R implementations: the EM algorithm for logistic and linear regressions with the package \pkg{misaem} \citep{jiang2020logistic} which uses the Stochastic Approximation of EM algorithm  \citep[SAEM][]{delyon1999convergence} and multiple imputation with the package \pkg{mice}. Both strategies are valid under the MAR missing data mechanism.
The workflow performs the estimation on a simulated dataset, but the dataset can be replaced with any custom dataset that the user believes satisfies the assumptions about the missing data mechanism and distribution of covariates.

The \pkg{misaem} package facilitates estimation of parameters of linear and logistic regression models from incomplete data, and also provides valid estimates of these parameters' variances. 
The functions \texttt{miss.lm},  \texttt{miss.glm} resemble the standard \texttt{lm} and \texttt{glm} functions both in terms of their signature and output. 

The rationale behind the popular multiple imputation approach is to create $M>1$ complete datasets by imputing the missing values with \samp{plausible} values, and then to estimate a parameter of interest $\theta$ on each of the imputed datasets. The multiple estimations of $\theta$ and their variability reflect the uncertainty due to the unknown missing values. The parameter estimation is performed by applying the analytic method used had, the data been complete. This provides an estimate of the parameter $\theta$ and an estimate of the corresponding variance, for each imputed dataset. These quantities are finally \samp{pooled} by using specific rules named \samp{Rubin’s rules} \citep{rubin2004multiple}, leading to a final point estimate, with a corresponding estimation of its variance that takes into account the uncertainty due to missing values.

In the corresponding workflow, we compare this method to the previous EM algorithm and provide the basic lines of code required to estimate parameters of linear or logistic regression models with incomplete covariate data.

For a additional example of how to estimate regression parameters we refer to the \href{https://rmisstastic.netlify.app/tutorials/josse_bookdown_dataanalysismissingr_2020}{ tutorial}\footnote{\url{https://rmisstastic.netlify.app/tutorials/josse_bookdown_dataanalysismissingr_2020}} on handling missing values in {R} by Julie Josse: it walks through a complete analysis, covering visualization of missing data patterns, data visualization, dimensionality reduction of incomplete data, and regression, in the presence of missing data.  

\subsection{How to predict in the presence of missing values?}
\label{sec:howtopred}

As mentioned in the introduction, methods to deal with missing values are not the same when the aim is to estimate parameters or to predict a target variable. \citet{josse_etal_2019} study the problem of supervised learning with missing values, i.e., when the aim is to predict an outcome $y$, from incomplete covariates in $X$. Note that contrary to the estimation setting, supervised learning involves training and test sets and both may have missing values. \citet{josse_etal_2019} recommend to impute the training set and the test set with a same constant, such as the mean, and then to apply a universally consistent learner, i.e., a very powerful learner, such as gradient boosting,  able to learn or fit any function. When  forests-based methods are used to do prediction, another method is available, the Missing Incorporated in Attributes (MIA) method \citep{twala2008good}. Note that constant imputation or MIA are recommended asymptotically but when having limited data in the prediction setting, other imputation methods can outperform these asymptotically consistent methods \citep{josse_etal_2019}. This is explored in the following workflows. The different methods are compared in terms of quality of the prediction of the outcome (AUC for a binary outcome and MSE for a continuous outcome).

\paragraph*{In {R}} The {R} \href{https://rmisstastic.netlify.app/how-to/external/how_to_predict_in_r}{workflow}\footnote{\url{https://rmisstastic.netlify.app/how-to/external/how_to_predict_in_r}} assesses a popular strategy (two-step strategy) which involves independently imputing the training and testing sets using the same imputation method. These datasets are then treated as being complete data, and regular learning algorithms are applied to predict some target variable.\footnote{This workflow has been written by an external contributor of the website, Katarzyna Woźnica (PhD student at the Warsaw University of Technology, Poland).} Several imputation methods are compared, such as \pkg{mice}, \pkg{missForest}, \pkg{softImpute}, and mean imputation. 
Note that, until recently, using the popular \pkg{mice} package for learning predictive models on incomplete data in {R} was hindered by the fact that it did not allow using the same imputation model for the training and test set. This has, however, been addressed with the argument \texttt{ignore} of the {R} function \texttt{mice}, the details of this recent extension can be found on GitHub.\footnote{\url{https://github.com/amices/mice/issues/32}}

\paragraph*{In {Python}} The {Python} \href{https://rmisstastic.netlify.app/how-to/python/predict_html/how\%20to\%20predict}{workflow}\footnote{\url{https://rmisstastic.netlify.app/how-to/python/predict_html/how\%20to\%20predict}} compares two strategies, where the aim is to predict a target variable and the covariates may contain missing values:
\begin{enumerate}
    \item The \textit{two-step} strategy consists of imputing the missing values both in the training and in the test set with a method like mean imputation or \texttt{IterativeImputer} of the \texttt{scikit-learn} library, and to apply usual learning algorithms (random forests, gradient boosting, linear regression) on the imputed dataset.
    This learning algorithm can be applied to the imputed dataset $\tilde{X}$ but also to a new variable made of the combination of the imputed covariates $\tilde{X}$ with the response pattern $R$: $[\tilde{X},R]$. 
    \item The \textit{one-step} strategy performs prediction using with learning methods adapted to the missing data without necessarily imputing them, such as the MIA method \citep{twala2008good}, which we have implemented in our notebook.
\end{enumerate}

We propose a function, \texttt{score\_pred}, which compares these strategies in terms of prediction performances by introducing missing values in complete covariates (\texttt{x\_comp}) under a specific missing data mechanism (\texttt{mecha} and a given percentage of missing values (\texttt{p}). The code for calling this function is given below, when the learning algorithm is the gradient boosting and 20\% of MCAR values are introduced.

\begin{example}
learner = HistGradientBoostingRegressor() 
p = 0.2
res = score_pred(x_comp=X, y = y, learner=learner , p=p, nbsim=10, mecha="MCAR")
\end{example}

\begin{figure}
\centering
\includegraphics[width=1\textwidth]{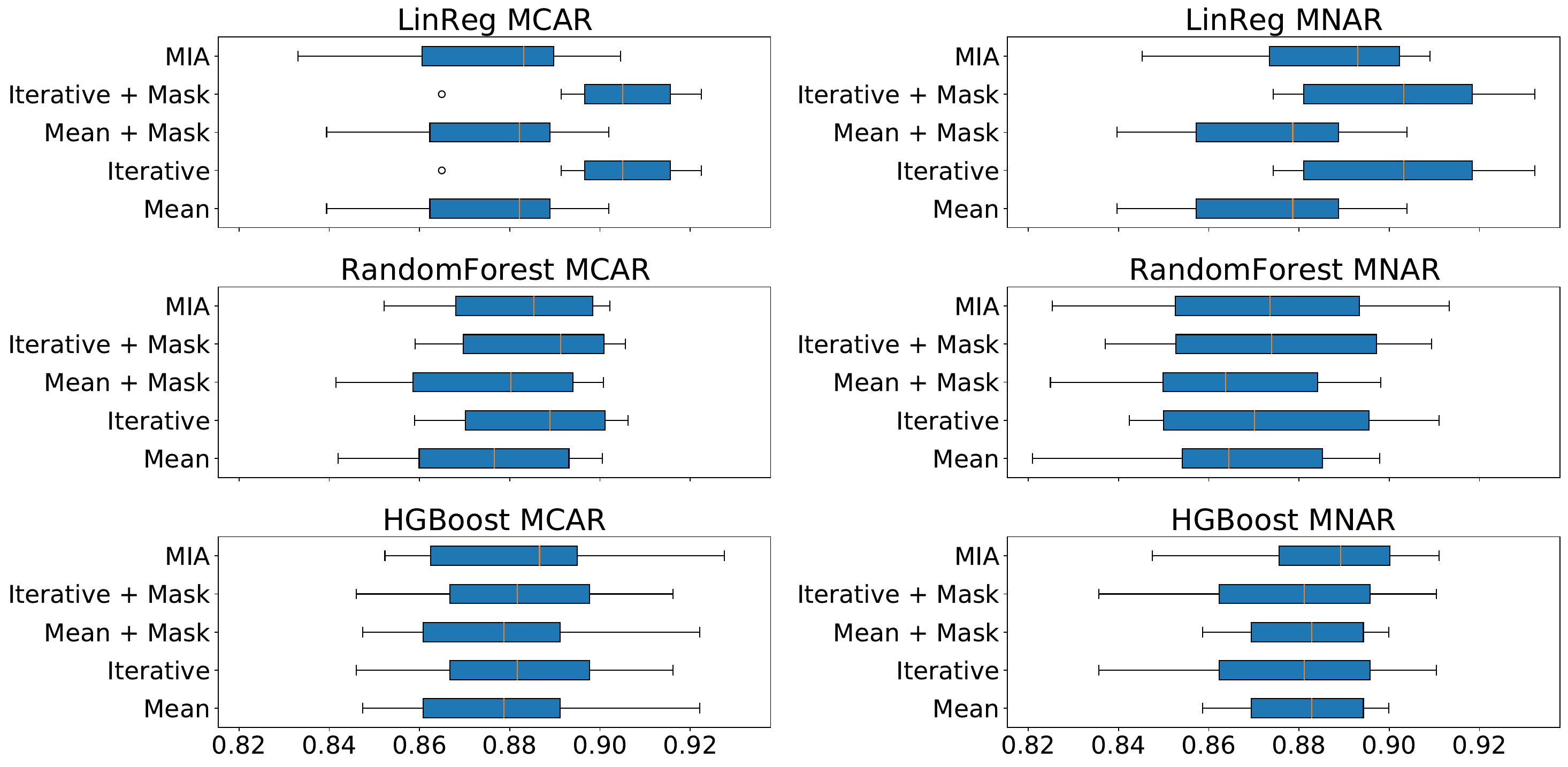}
\caption{\label{fig:workflows2} Plot of the function \texttt{score\_pred} to compare different strategies when the aim is to predict in {Python}. 20\% of missing values are introduced in a simulated dataset using the MCAR mechanism or the MNAR mechanism. The covariates $X\in \mathbb{R}^{1000\times 3}$ are generated under a multivariate Gaussian distribution, the parameter of the regression $\beta \in \mathbb{R}^3$ follows a random uniform distribution. The outcome $y$ is generated according to a linear  model such that $y=X\beta+\epsilon,$ with $\epsilon$ representing Gaussian noise. The two-step strategies (\texttt{IterativeImputer} and the mean imputation) with or without adding a mask and the one-step strategy \textit{MIA} are compared in terms of prediction error, and several learners are performed (linear regression, random forests, gradient boosting). The closer the result is to $1$, the more accurate the prediction is ($1$ corresponds to perfect prediction, $0$ to the worst prediction).}
\end{figure}

The dataset is then split into a training set and a test set (75\% in the training set, 25\% in the test set) and the methods presented below are applied by considering a specific learning algorithm. The function then returns the prediction error on the test set, by comparing the ground truth (\code{y}) and the predicted outcome values on the test set for each simulation (i.e., each run for the generation of missing values). Figure \ref{fig:workflows2} shows the graphical output of this function called for different learning algorithms (linear regression, random forests and gradient boosting) and for different missing data mechanisms (20\% MCAR and MNAR, see the section on how to generate missing values). 
When the learner is the linear regression, the two-step methods with added mask, both for the MCAR mechanism and the MNAR mechanism, perform well. Since the simulated dataset is generated using a linear regression, the linear regression is expected to give better results than the other learners. In addition, for the MNAR mechanism, the one-step strategy \textit{MIA} (especially when the gradient boosting is performed) appears to be a good choice. 

Another function is specifically designed to handle datasets which already contain missing values. The second part of the appendix shows a concrete example of this notebook on a real dataset. 

This concludes the overview of the workflows developed in this project. We invite other practitioners and researchers to use and extend these methods. Overall, we hope that by creating and sharing methods, new methods can be more easily developed and easily compared and evaluated.

\section{Perspectives and future extensions}

By providing a platform and community to discuss missing data, software, approaches and workflows, the sharing of expertise on missing data can hopefully be improved and extended more easily.

\label{sec:conc}

\subsection{Towards uniformization and reproducibility}

One way to promote and encourage practitioners and researchers in their work with missing values is to provide benchmarks and workflows around missing data. As has been shown in data competitions, community involvement produces many creative solutions and discussions that move the field forward, and challenge existing strategies.
We will continue to work on our workflows and related source code. In doing so, we hope to encourage users to continue to test new methods and present results in a clear and reproducible manner.
In addition, we plan to propose two types of data challenges: 1) imputation and estimation, and 2) analysis workflows.
For the first part of the challenge, the objective is to find the best imputation or estimation strategy. The community will be given a dataset with missing values, for which there is actually a hidden copy of the real values. The community will then get the task of creating imputed values, which are assessed against the original dataset with complete values, to determine which imputation is best. This is similar in spirit to the Netflix prize \citep{Bennett2007} and the M4 challenge in the time series domain \citep{Makridakis2018}. This benchmarking could be extended to other areas, such as parameter estimation, and predictive modeling with missing data.
Analysis workflows could form another community challenge, assessed in a similar way to existing \samp{datathon} events where entries are assessed by an expert panel.
Here the challenge could be to develop workflows and data visualizations from complex data. The data could have challenging features, and be combined from various data sources with complex structure, such as data with several types of missingness, images, text, data, longitudinal data, and time series.

\subsection{Future extensions}

Possible enhancements that could be added in future releases of the platform, for which we welcome suggestions and contributions, are the following: a workflow with a focus on MNAR data and different solutions that can handle such data (as diversity of existing solutions is large, such a unified workflow will be a consequential contribution); for more applied users, a comparison of computation times of different methods, benchmarked on various types of data. 
Another problem that is becoming more common is missing values in data integration. Indeed, questions such as \textit{what do I do when I have clinical data from multiple centers with different mechanisms of missing values or with systematically missing values in certain data?} or \textit{what do I do when I have time series and missing values in one of the groups of variables?} would be also worth addressing in additional workflows.

\subsection{Participation and interaction}

This platform is aimed to offer a venue for the community, in the sense that we welcome
every comment and question, encourage submissions of new works, theoretical or
practical, either through the provided contact form or directly via the GitHub
project repository. We
have already received useful feedback and several external contributions, organized several remote
calls and working sessions at statistics conferences. We are
planning on regularly relaunching calls for new material for the platform, for example through the R consortium
blog\footnote{\url{https://www.r-consortium.org/news/blog}},
R-bloggers\footnote{\url{https://www.r-bloggers.com/}} and social media
platforms. We also intend to use these channels to communicate more generally
about the platform and the topic of missing values.

In order for the platform to be a reference to the community, it must 
provide regularly updated, user-friendly content. To achieve this goal, it is important to propose sustainable and accessible solutions for the maintenance of the \samp{R-miss-tastic} platform. We hope that the well documented source code of the platform facilitates external contributions and community feedback on this project.

In conclusion, the aim of this platform is to go beyond mere community participation, namely to seed meaningful community interactions, and to offer a hub of communication among groups that rarely exchange, both within, and between academia and industry. 

\section{Acknowledgements}
This work has partially been funded by the R Consortium, Inc. We would like to thank Steffen \textsc{Moritz} and Fran\c{c}ois \textsc{Husson} for their active support and feedback, all contributors who have generously made their course and tutorial materials available, as well as the contributors to the workflows in {R} and {Python} code.


\newpage

\begin{appendix}

\section{Appendix}

\subsection{Tutorial for imputing missing values in {R}} \label{app:tutoR}

The goal of this tutorial is to give practical details on the \href{https://rmisstastic.netlify.app/how-to/impute/missimp}{{R}-workflow} entitled \textit{How to impute missing values?}\footnote{This tutorial is only an example of use but more practical details are given in the original workflow.}. In this workflow, users can compare the most popular methods to impute missing values in {R} on simulated or real datasets. 

We illustrate this workflow by considering a small dataset called \textit{decathlon}, that contains athletes' performance during two sporting events (41 rows, 13 columns\footnote{We do not consider the last variable in this part, which is categorical. Some imputation methods do not handle mixed data.}). It is available in the {R}-package \pkg{FactoMineR} \citep{le2008factominer}.

\begin{example}
library(FactoMineR)
data(decathlon)
head(decathlon[,1:4]) # four first columns of the  dataset
         100m Long.jump Shot.put High.jump
SEBRLE  11.04      7.58    14.83      2.07
CLAY    10.76      7.40    14.26      1.86
KARPOV  11.02      7.30    14.77      2.04
BERNARD 11.02      7.23    14.25      1.92
YURKOV  11.34      7.09    15.19      2.10
WARNERS 11.11      7.60    14.31      1.98
\end{example}

If we have collected similar data, e.g., described by the same variables but for new athletes, that contain missing values, practitioners may want to know how to impute such a dataset. To address this question, we can introduce missing values under different mechanisms (MCAR, MAR or MNAR) and with different percentages of missing values (here we compare 20\% and 50\%) in the complete dataset and compare some imputation methods in terms of mean squared error (MSE), i.e., the error committed by the imputation of the missing values. Missing values are introduced in all covariates. The function \texttt{how\_to\_impute} can be used to compare the imputation methods described in the section on how to impute missing values (\pkg{missMDA}, \pkg{mice}, \pkg{missForest}, \pkg{softImpute} and the imputation by the mean) using different percentages and types of missing values given in two lists by the users. More particularly, the arguments are the following ones: the complete dataset where the missing values will be introduced (\code{X}), a list containing the different percentage of missing values (\code{perc.list}), a list containing the different missing-data mechanisms (\code{mecha.list}) and the number of simulations performed (\code{nbsim}). 
Note that for \pkg{missMDA}, the number of components in the PCA used to predict the missing entries is estimated using a cross-validation with the function \texttt{estim\_ncpPCA}. For \pkg{softImpute}, we use a cross-validation to choose the regularization parameter (coded for the purpose of the notebook).
This function returns a table with the mean of the MSEs over the simulations for the different methods and for the different missing data settings (20\% MCAR values, 50\% MCAR values, 20\% MAR values, 50\% MAR values, 20\% MNAR values, 50\% MNAR values).

\begin{example}
perc.list <- c(0.2,0.5)
mecha.list <- c("MCAR", "MAR", "MNAR")
res <- how_to_impute(X=decat_sc, perc.list=perc.list,mecha.list=mecha.list,nbsim=10)
res

          0.2 MCAR  0.5 MCAR   0.2 MAR   0.5 MAR  0.2 MNAR  0.5 MNAR
X.pca    0.8822782 1.0537611 0.9394561 1.0873315 0.9876867 1.1026891
X.forest 0.8820789 0.9577351 0.9403659 1.0526915 0.9940827 1.0809478
X.mice   0.8610320 1.0372518 0.9559042 1.0948981 0.9887239 1.1271581
X.soft   0.7935545 0.8865989 0.8721907 0.9556373 0.8951239 0.9692859
X.mean   1.0177192 1.0306089 1.0972080 1.0770715 1.1342289 1.1077467
\end{example}

\begin{figure}
\begin{subfigure}[b]{1\textwidth}
\centering
\includegraphics[width=0.7\textwidth]{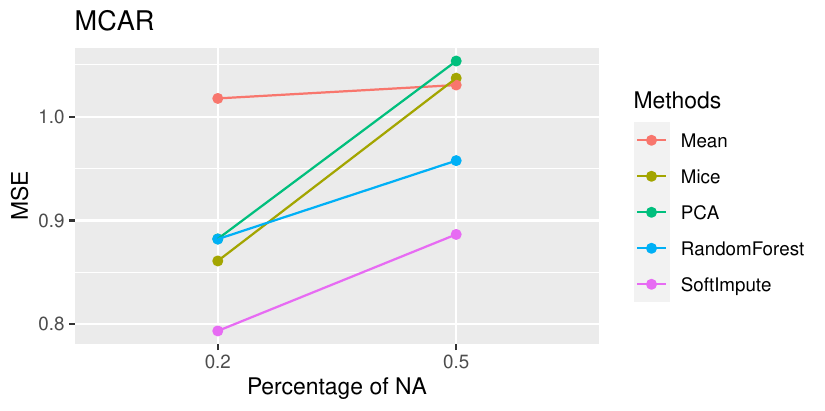}
\end{subfigure}
\begin{subfigure}[b]{1\textwidth}
\centering
\includegraphics[width=0.7\textwidth]{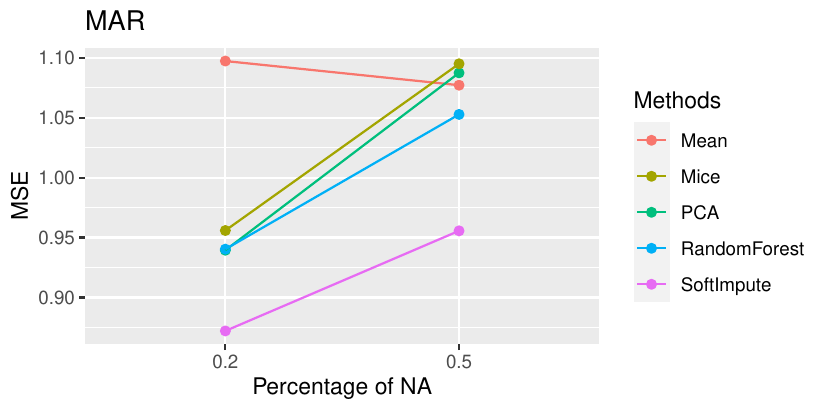}
\end{subfigure}
\begin{subfigure}[b]{1\textwidth}
\centering
\includegraphics[width=0.7\textwidth]{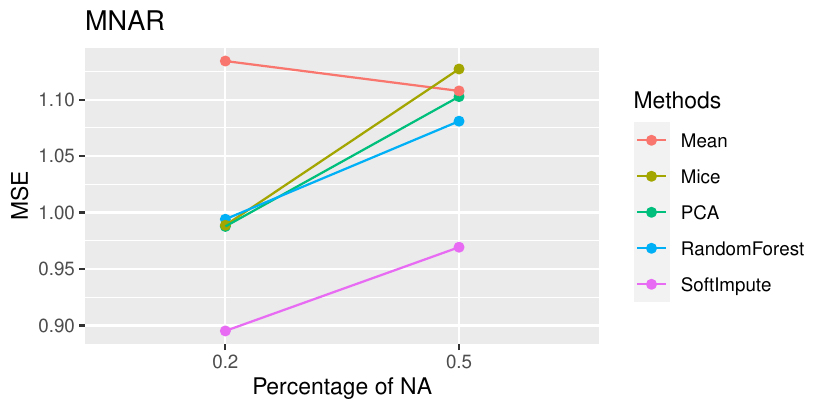}
\end{subfigure}
\caption{\label{fig:howtoimpute_R} Graphical outputs of the {R} function \texttt{how\_to\_impute}. The methods \pkg{mice}, \pkg{missForest}, \pkg{softImpute} and \pkg{missMDA} are compared with the naive imputation by the mean for several percentages of missing values (20\%, 50\%). The mean of the MSEs computed for several generations of missing values are given. The results are shown for different mechanisms (MCAR, MAR, MNAR).}
\end{figure}

With this result in hand, we can easily visualize some of the results. Figure \ref{fig:howtoimpute_R} shows the associated graphics, for each of the missing-data mechanism. The code to obtain these graphics is given below. 

\begin{example}
plotdf <- do.call(c, res)
plotdf <- as.data.frame(plotdf)
names(plotdf) <- 'mse'
n_perc.list <- length(perc.list) 
n_mecha.list <- length(mecha.list)
methods.list <- c("PCA", "RandomForest",  "Mice", "SoftImpute", "Mean")
meth <- rep(methods.list, n_perc.list * n_mecha.list)
plotdf <- cbind(plotdf, meth)
perc <- rep(rep(as.character(perc.list), each = 5),length(mecha.list))
plotdf <- cbind(plotdf, perc)
mecha <- rep(mecha.list, each = 5 * length(perc.list))
plotdf <- cbind(plotdf, mecha)

# For MCAR data 
ggplot(plotdf[plotdf['mecha'] == "MCAR",]) 
+ geom_point(aes(x = perc, y = mse, color = meth), size = 2) 
+ ylab("MSE") + xlab("Percentage of NA") 
+ geom_path(aes(x = perc, y = mse, color = meth, group = meth)) 
+ ggtitle("MNAR") + labs(color = "Methods") + theme(text = element_text(size = 20)) 

# For MAR data 
ggplot(plotdf[plotdf['mecha'] == "MCAR",]) 
+ geom_point(aes(x = perc, y = mse, color = meth), size = 2) 
+ ylab("MSE") + xlab("Percentage of NA") 
+ geom_path(aes(x = perc, y = mse, color = meth, group = meth)) 
+ ggtitle("MNAR") + labs(color = "Methods") + theme(text = element_text(size = 20)) 


# For MNAR data 
ggplot(plotdf[plotdf['mecha'] == "MCAR",]) 
+ geom_point(aes(x = perc, y = mse, color = meth), size = 2) 
+ ylab("MSE") + xlab("Percentage of NA") 
+ geom_path(aes(x = perc, y = mse, color = meth, group = meth)) 
+ ggtitle("MNAR") + labs(color = "Methods") + theme(text = element_text(size = 20)) 
\end{example}

For this dataset and these missing data settings, \pkg{softImpute} appears to be the best imputation method. 

\subsection{Tutorial for predicting in presence of missing values in Python} \label{app:tutopython}

The goal of this tutorial is to give practical details on the \href{https://rmisstastic.netlify.app/how-to/python/predict_html/how\%20to\%20predict}{{Python}-workflow} entitled \textit{How to predict with missing values?}\footnote{This tutorial is only an example of use but more practical details are given in the original workflow.}. In this workflow, users can compare methods in {Python} to predict a target variable when the covariates contain missing values. 

We consider the dataset called \textit{california\_housing} (20640 rows, 9 columns). The target variable is the median house value for California districts and the covariates provide information (latitude, longitude, number of people in the district...) on the different districts. 

If we know that new observations will contain missing values, an interesting question is how to predict the target variable in presence of covariates with missing values. To answer this, we can impute missing values in the covariates and compare methods which handle them and predict the target variable. 

First, we generate missing values in the covariates of the dataset \textit{california\_housing}, using the function \texttt{produce\_NA}. The three main arguments are the initial dataset (\texttt{X}) in which missing values are introduced using a given missing data mechanism (\texttt{mecha}) and a given percentage of missing values (\texttt{p\_miss}). In the following example, we introduce 20\% MCAR values. 

\begin{example}
XproduceNA_MCAR = produce_NA(X = x_comp, p_miss = 0.2, mecha = "MCAR")
x_MCAR = XproduceNA_MCAR['X_incomp'].numpy()
\end{example}

\begin{figure}
\includegraphics[width=0.32\textwidth]{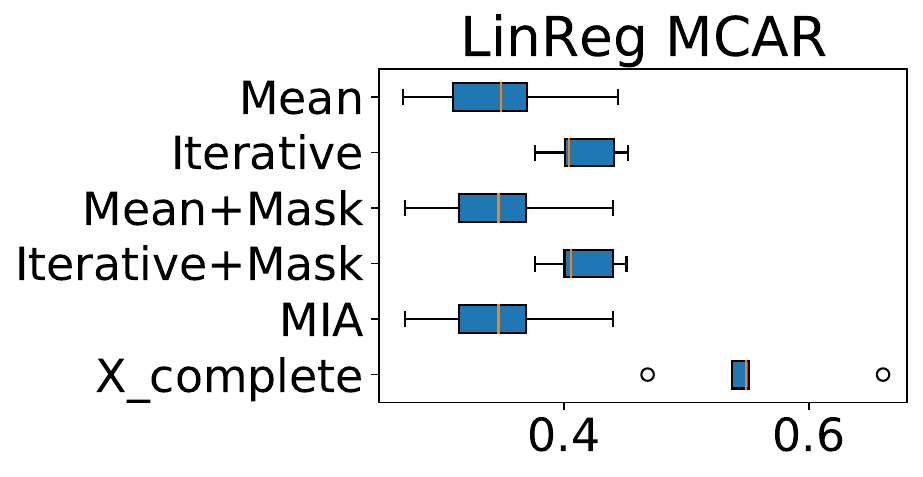}
\includegraphics[width=0.32\textwidth]{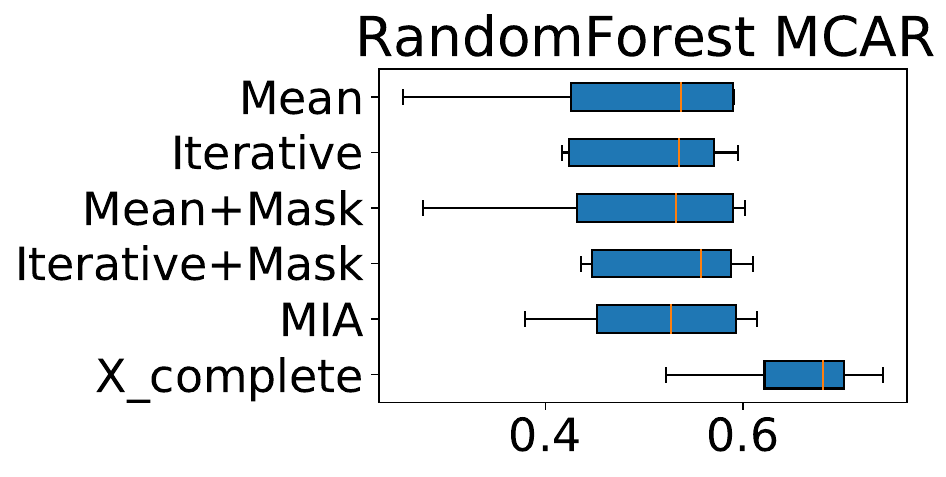}
\includegraphics[width=0.32\textwidth]{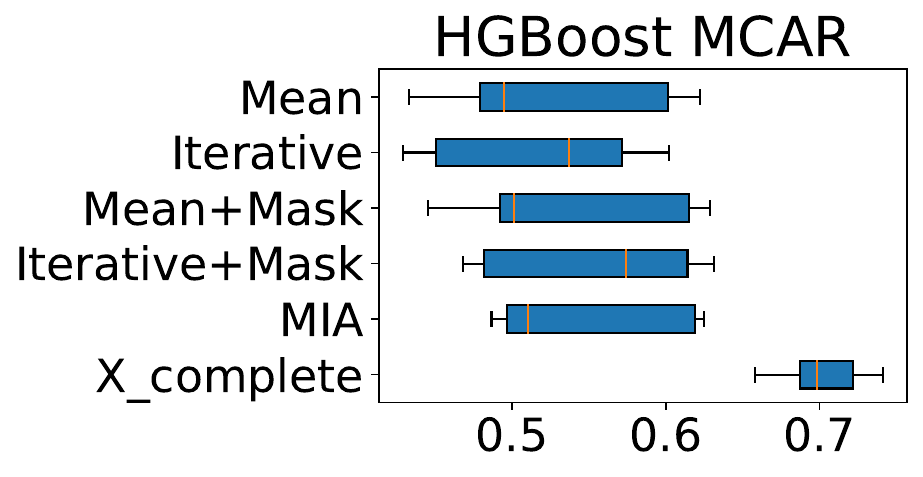}
\caption{\label{fig:howtopredict_Python} Graphical outputs of the {Python} function \texttt{plot\_score\_realdatasets}. The x-axis indicates the MSE. The two-steps methods considering the imputation by the mean (Mean), IterativeImputer (Iterative) with or without adding the mask and the one-step method MIA are compared with the case without missing values (X\_complete). Note that the mask is the binary matrix which indicates where are the missing values. When we add the mask, we consider then an augmented matrix, with the initial matrix and the mask. The learner are the linear regression (left graphic), the random forests (middle graphic) and gradient boosting (right graphic).}
\end{figure}

To predict, we consider two strategies presented in the section on how to predict in the presence of missing values: (i) the \textit{two-step} strategy which consists of imputing missing values and applying classical methods on the completed data sets to predict, and (ii) the \textit{one-step} strategy which predicts using methods adapted to the missing values without necessarily imputing them. The code below allows comparison of different prediction strategies nested in a two-step or a one-step strategy. To do so, we use the function \texttt{plot\_score\_realdatasets}, which handles datasets already containing missing values. Figure \ref{fig:howtopredict_Python} shows the graphical output of this function called for different learning algorithms. When the learner is the linear regression, the two-step methods with added mask perform well. Indeed, since the simulated dataset is generated considering a linear regression, the linear regression is expected to give better results than the other learners. 

The code for the function \texttt{plot\_score\_realdatasets} is given below. The main arguments are the dataset (\texttt{X}), the outcome variable (\texttt{y}) and the learning algorithm to use (\texttt{learner}). 

\begin{example}
learners = {'LinReg': LinearRegression()
            'RandomForest': RandomForestRegressor(),
            'HGBoost': HistGradientBoostingRegressor()}
for learner_name, learner in learners.items():
    plt.figure(figsize=(10,10))
    for ii, (X, X_name) in enumerate(zip([x_MCAR], ['MCAR'])):
        plt.subplot(3, 2, ii+1)
        plot_score_realdatasets(X, y, learner, learner_name + ' ' +X_name,x_comp)
\end{example}

\begin{table}[]
\small
\begin{tabular}{c|c|c|c}
\textbf{R implementation}                                                                                 & \textbf{Scope}                                                                                                                              & \textbf{Python counterpart}                                                                                           & \textbf{Differences (if any)}                                                                                                                                                                                                                                                                                                                                                                                \\ \hline
\begin{tabular}[c]{@{}c@{}}imputeMean\\ (implemented in \\ R-miss-tastic \\ workflow)\end{tabular} & \begin{tabular}[c]{@{}c@{}}Impute missing\\ values\\ of each variable \\ by their means.\end{tabular}                                 & \begin{tabular}[c]{@{}c@{}}module \\ sklearn.impute, \\ \texttt{SimpleImputer} \\ with \texttt{strategy='mean'}\end{tabular}           &                                                                                                                                                                                                                                                                                                                                                                                                       \\ \hline
\pkg{softImpute}                                                                                        & \begin{tabular}[c]{@{}c@{}}Impute missing \\ values \\ using a low-rank \\ completion \\ with nuclear norm \\ penalities\end{tabular} & \begin{tabular}[c]{@{}c@{}}function softImpute \\ (implemented\\ in R-miss-tastic \\ workflow)\end{tabular}          & \begin{tabular}[c]{@{}c@{}}The R package is better \\ optimized than our \\ Python version. \\ The R implementation \\ also has different \\ optimization \\ algorithms implemented \\ (itervative SVD, \\ iterative ALS).\end{tabular}                                                                                                                                                               \\ \hline
\pkg{mice }                                                                                             & \begin{tabular}[c]{@{}c@{}}Give multivariate \\ imputations by \\ chained equations\end{tabular}                                      & \begin{tabular}[c]{@{}c@{}}module \\ sklearn.impute\\ \texttt{IterativeImputer} \\ with \\ \texttt{BayesianRidge}\end{tabular}         & \begin{tabular}[c]{@{}c@{}}The Python module uses \\ iterative chained equations. \\ However, it differs from \\ the mice package, \\ because it uses a ridge iterate \\ and it returns by default \\ a single imputation. \\ Note that the argument \\ \texttt{sample\_posterior=True} allows \\ to get stochastic imputations, \\ and not multiple imputations, \\ as the R function mice does.\end{tabular} \\ \hline
\pkg{missForest}                                                                                     & \begin{tabular}[c]{@{}c@{}}Impute missing \\ values \\ using random \\ forests\end{tabular}                                           & \begin{tabular}[c]{@{}c@{}}module \\ sklearn.impute, \\ \texttt{IterativeImputer} \\ with \\ \texttt{ExtraTreesRegressor}\end{tabular} & \begin{tabular}[c]{@{}c@{}}The Python \\ implementation\\ does not exactly use random \\ forests with CART trees,\\  but forests with trees \\ which choose a random \\ split (instead of the best\\ split per feature).\end{tabular}                                                                                                                                                                 \\ \hline
\pkg{missMDA}                                                                                        & \begin{tabular}[c]{@{}c@{}}Impute missing values \\ using a low-rank matrix \\ completion with penality\end{tabular}                  &                                                                                                                      &                                                                                                                                                                                                                                                                                                                                                                                                       \\ 
\end{tabular}
\caption{\label{tab:compar}Comparison table on the difference of the scope between R and Python packages used in the R-miss-tastic workflows.}
\end{table}

\end{appendix}

\bibliography{r-miss-tastic}

\address{Imke Mayer \\
  Institute of Public Health, Charité – Universitätsmedizin Berlin \\
  \email{imke.mayer@charite.de}}

\address{Aude Sportisse\\
  3iA Côte d'Azur, Centre Inria d'Université Côté d'Azur
  \email{aude.sportisse@inria.fr}
}

\address{Nicholas Tierney \\
Department of Econometrics and Business Statistics, Monash University \\
\email{nicholas.tierney@gmail.com}
}

\address{Nathalie Vialaneix \\
MIAT, Universit\'e de Toulouse, INRA \\
\email{nathalie.vialaneix@inrae.fr}
}

\address{Julie Josse \\
PreMeDICaL - Precision Medicine by Data Integration and Causal Learning, Inria Sophia Antipolis \\
\email{julie.josse@inria.fr}
}

\end{article}

\end{document}